\documentclass[aps,prl,preprint,superscriptaddress]{revtex4-1}
\usepackage{graphicx}
\usepackage{dcolumn}
\usepackage{bm}
\usepackage{amssymb}
\usepackage{latexsym}
\usepackage{verbatim}
\usepackage[colorlinks=true, linkcolor=red, citecolor=blue]{hyperref}
\begin{document}

\title{Superconducting-contact-induced resistance-anomalies in the 3D topological insulator
Bi$_2$Te$_3$}
\author{Zhuo Wang}
\author{Tianyu Ye}
\author{R. G. Mani}
\affiliation{Department of Physics and Astronomy, Georgia State
University, Atlanta, GA 30303.}

\date{\today}

\begin{abstract}
This study examines the magnetotransport response observed in flakes
of the $3$D topological insulator Bi$_2$Te$_3$, including indium
superconducting electrodes, and demonstrates two critical
transitions in the magnetoresistive response with decreasing
temperatures below $T= 3.4$ K. The first transition is attributed to
superconductivity in the indium electrodes, and the second
transition, with a critical field exceeding the transition field of
indium, is attributed to a proximity effect at the 2D planar
interface of this hybrid TI/superconductor structure.

\end{abstract}

\maketitle

Topological insulators (TI) are 2- and 3-dimensional electronic
materials that behave as a normal insulator in the bulk but conduct
electricity in edge or surface states, respectively, that arise from
a bulk-boundary correspondence, and are topologically protected by
time-reversal symmetry.\cite{Hasan2010, Qi2011, Ando2013} TI are
characterized by a strong spin-orbit coupling that provides a
locking between spin and linear momentum, and the helical spin
polarization as gapless surface states exhibit a linear dispersion
relation, i.e., "Dirac cones," characteristic of relativistic
massless particles. The $CdTe/HgTe/CdTe$ quantum well system was the
first theoretically predicted\cite{Bernevig2006} and experimentally
demonstrated\cite{Konig2007} 2D TI, followed by the
prediction\cite{Liu2008} and experimental
confirmation\cite{Knez2011} of the technological significant
$AlSb/InAs/GaSb/AlSb$ quantum well system as another 2D TI.
$Bi_{1-x}Sb_{x}$ was the first theoretically predicted- and
experimentally identified-3D TI.\cite{Fu_Kane2007, Hsieh2008} Since
then, a large number of other systems with a band-gap $E_{0} \le 0.3~eV$ at $300$ K including $Bi_{2}Se_{3}$,\cite{Xia2009}
$Sb_{2}Te_{3}$,\cite{Jiang2012} and
$Bi_2Te_3$,\cite{Chen2009,Hsieh2009} have joined the class of 3D TI.
Remarkably, among these systems, $Bi_{2}Se_{3}$ has exhibited
superconductivity with a critical transition temperature of $T_c =
3.8$ K, when copper is intercalated between the adjacent quintuple
layers.\cite{Hor2010,Sasaki2011} Further, the application of
pressure in the range of 3-6 GPa in $Bi_2Te_3$ has been shown to
induce superconductivity with $T_c \approx 3$ K and $B_c(1.8~$K$)
\approx 0.2$ T.\cite{Zhang2011}

The proximity effect from an ordinary s-wave superconductor in the
TI  leads to exotic $p_{x} + ip_{y}$ superconductivity capable of
hosting Majorana fermions, and this prediction has heightened
interest in experimental studies of superconductor-topological
insulator hybrid devices.\cite{Fu_Kane2008, Nayak2008}  Thus, many
have examined TI nanowire-superconductor devices in search of
Majorana modes at the ends of one dimensional devices.\cite{
Mourik2012, Das2012, Rokhinson2012, Deng2012, Chang2013, Finck2013}
Nevertheless, a two-dimensional experimental setting  for the study
of Majorana modes is desirable. In Ref. \cite{Fu_Kane2008},
superconductivity is induced in the surface state of the TI by the
proximity effect produced by an ordinary s-wave superconductor, and
the Majorana mode is a vortex bound state at the interface of the
two materials.  Recently, there has been interest in  whether---in systems now called "topological metals," which include 3D TI
materials such as $Bi_2Te_3$ and $Bi_{2}Se_{3}$ that have bulk
states coexisting with surface states at the Fermi energy---it is
possible for the proximity effect to extend from the superconductor
covered surfaces to uncovered surfaces of the 3D TI\cite{Wang2011}
and whether the uncovered surfaces of the 3D TI can thus host
Majorana modes.\cite{Lee2014}  In moving towards exploring such
possibilities, we have examined the magnetotransport response in the
$3$D topological insulator $Bi_2Te_3$ in the presence of (indium)
topside superconducting electrodes.\cite{Mani1992, Ghenim1993,
Zhangd2011, Yang2012, Veldhorst2012, Qu2012} Remarkably, the
experimental results showed two critical transitions in the
magnetoresistive response below $T= 4.2$ K. While the first
transition is associated with a shunting effect by the contacts, the
second anomalous contribution, which indicates $T_c \approx 3.1$ K
and $B_c(1.5~$K$) \approx 0.2$ T, is attributed here to a
generalized proximity effect in the TI.

$Bi_2Te_3$ flakes ($\approx 25 \mu m$ thick) were mechanically
exfoliated using a scotch tape method from single  crystals of
$Bi_2Te_3$ and transferred onto $Si/SiO_2$ substrates. The
$Bi_2Te_3$ flakes were approximately shaped like rectangles with the
length-to-width ratio $L/W \approx 2$. Superconductor/topological
insulator junctions were realized by directly pressing indium onto
the surface of the $Bi_2Te_3$ flake in a Hall bar configuration; see
supplementary material for the images of the
samples.\cite{supplementary} Here, indium is a s-wave superconductor
with the critical temperature $T_c = 3.41$
K.\cite{SuperconductivityTc} Electrical measurements were carried
out using the standard four-terminal low-frequency lock-in techniques,
and all the reported resistances are four-terminal resistances. The
\textit{ac} was applied along the long axis of the Hall bar,
through electrodes at the two ends, and the magnetic field was
applied perpendicular to the Hall bar surface, as usual. Since the
flakes are not so thin, the applied current is carried by both the
specimen bulk and the specimen surfaces. Further, since the
proximity effect should be restricted to the near surface regions,
one expects to observe just resistance corrections due to the
proximity effect in the measured four terminal resistances.  That
is, one does not expect the four terminal resistance to vanish even
with the onset of superrconductivity in the contacts and the
proximity effect in the surfaces near the contacts. The specimens
were immersed in pumped liquid Helium, and the temperature was varied in the range $1.6 \le T \le 4.2$ K by controlling the vapor pressure of
liquid helium within a $^4$He cryostat. Hall effect measurements
indicated the carrier concentration $n \approx 10^{19} cm^{-3}$
in these samples.


Fig. 1a shows a color plot of the normalized magnetoresistance,
$\Delta R/R$ as a function of both the magnetic field, $B$, and the
temperature, $T$, for sample 1. Here, $\Delta
R/R=[R(B)-R(B_N)]/R(B_N)$, where $B_N=-0.3$ T. The color plot (Fig.
1(a)) shows a set of horizontal and vertical lines, which correspond
to constant $B$- and $T$-cross-sections, which are exhibited in
Fig. 1(b) and Fig. 1(c), respectively. In the $T=3.6$ K
cross-section, shown in light green (see Fig. 1(b)), sample $1$ shows a
featureless $\Delta R/R$, with a small positive magnetoresistance of
$0.3$ T.\cite{supplementary} However, at $3.4$ K, which is
slightly below the superconducting transition temperature of indium,
$\Delta R/R$ exhibits a sharp and narrow dip near zero magnetic
field. As the temperature is reduced further, this dip in $\Delta
R/R$ rapidly becomes deeper until $T = 3.1$ K, as its width along
the $B$-axis, denoted here as $\Delta B_1$, increases. A close
examination of $\Delta R/R$ suggests the emergence of a second
$\Delta R/R$ contribution for $T \leq 3.0$ K in this specimen. The
width but not the depth of this second contribution to $\Delta R/R$
grows rapidly with decreasing $T$. At $T=1.7$ K, the critical
magnetic field is denoted as $B_{c2} (1.7~$K$)= \Delta B_2/2 \approx
0.18$ T. Fig.1(c) depicts the magnitude of $\Delta R/R$ vs. $T$
at the constant-$B$ cross-sections shown in Fig. 1(a) . This figure
shows that, with decreasing temperatures, the $\Delta R/R$ drops
within a small $\Delta T$ interval and then plateaus. The $\Delta T$
interval where the drop is observed depends upon $B$. The
temperature dependence of the critical fields associated with the
two contributions, i.e., $B_{c1}$ and $B_{c2}$, are shown in Fig.1
(d). Here, the critical fields were determined by examining the
first derivative of the magnetoresistance data and identifying the
magnetic fields where the slope, $dR_{xx}/dB$, exhibits a
substantial change. The results of Fig. 1(d) indicate an
approximately linear relationship between the critical fields and
temperature for the two contributions to $\Delta R/R$.

\begin{figure}[t]
\centering \noindent
\includegraphics[width=7cm]{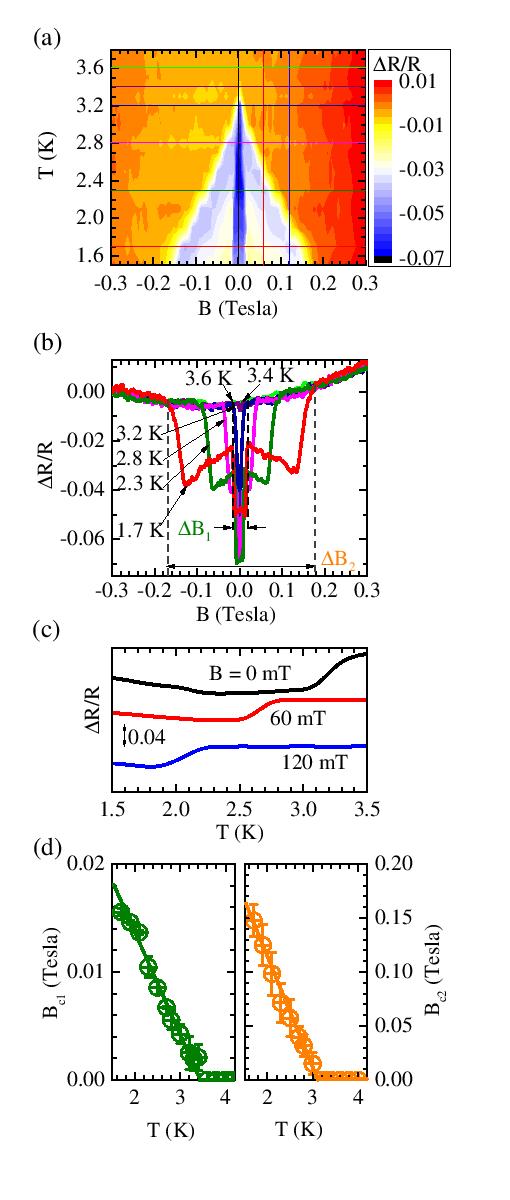}
\caption{\label{fig:Fig2} {\textbf{Transport in Bi$_2$Te$_3$ flakes including indium contacts.} (a) A 2D color plot of
the normalized magnetoresistance, $\Delta R/R$, is shown vs. the
magnetic field, $B$, and the temperature, $T$, for a $Bi_2Te_3$
specimen, sample $1$. (b) This panel exhibits the normalized
magnetoresistance~vs. $B$, at the $T$-cross-sections indicated in
panel (a). (c) This panel shows the temperature-dependence of
$\Delta R/R$ at zero magnetic field. (d) The critical fields $B_{c1}$ and $B_{c2}$ are plotted on the left and right, respectively, vs. $T$.  Data
are shown as symbols.  Lines represent the trend.}}
\end{figure}

The results for a second $Bi_2Te_3$ specimen are shown in Fig. 2. The
top panel in Fig. 2 shows a color plot of $\Delta R/R$ vs. $B$ and
$T$, with five horizontal- and three vertical-cross sections.  The data corresponding to the horizontal-cross sections are shown in Fig. 2(b). Here, $\Delta
R/R$  shows an initial negative magnetoresistance followed by a weak
positive magnetoresistance at $3.6$ K(see Fig.2(b)). Fig. 2 (b) also shows a rapid
dip in $\Delta R/R$ beginning at $T=3.4$ K in the vicinity of $B=0$.
A second resistance correction term is readily apparent by $2.8$ K,
and this term mainly gains width but not depth with  the decreasing temperature. At $T=2.0$ K, the critical magnetic field is denoted as
$B_{c2} (2.0~$K$)= \Delta B_2/2 \approx 0.12$ T in this specimen.
The temperature dependences of the critical fields associated with
the two contribution, i.e., $B_{c1}$ and $B_{c2}$, are shown in Fig.
2(d). The results indicate an approximately linear increase in the
critical fields with the decreasing temperatures.  The vertical
(constant magnetic field) cross sections indicated in Fig. 2(a) are
shown in Fig. 2(c). As in Fig. 1(c),  Fig.2(c)  shows that, with decreasing temperatures, the $\Delta R/R$ drops within a small
$\Delta T$ interval and then plateaus. The $T$-band where the drop
is observed depends again upon $B$.

\begin{figure}[t]
\centering \noindent
\includegraphics[width=7cm]{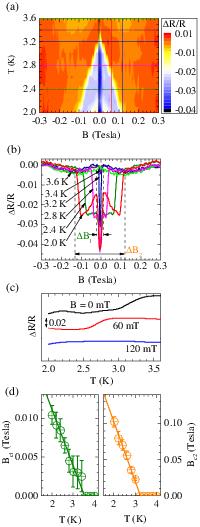}
\caption{\label{fig:Fig2} {\textbf{Transport in Bi$_2$Te$_3$ flakes including indium contacts.} (a) A 2D color plot of
the normalized magnetoresistance, $\Delta R/R$, is shown vs. the
magnetic field, $B$, and the temperature, $T$, for a $Bi_2Te_3$
specimen, sample $2$. (b) This graph exhibits the normalized
magnetoresistance~vs. $B$, at the $T$-cross-sections indicated in
the top panel. (c) This panel compares the temperature dependence
of the normalized resistance at $B=~0$ mT, $B=~60$ mT, and
$B=~120$ mT. (d) The critical fields $B_{c1}$ and $B_{c2}$ are plotted on the left and right, respectively, vs $T$  Data
are shown as symbols.  Lines represent the trend.}}
\end{figure}

\begin{figure}[t]
\centering \noindent
\includegraphics[width=7cm]{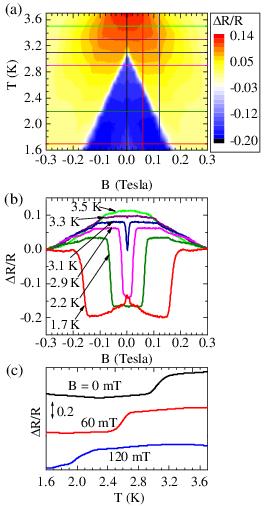}
\caption{\label{fig:Fig2} { \textbf{Transport in Bi$_2$Te$_3$ flakes including indium contacts.} (a) This figure shows a
2D color plot of normalized magnetoresistance, $\Delta R/R$, vs.
the magnetic field, $B$, and the temperature, $T$, for a
$Bi_2Te_3$ specimen, sample $3$. (b) The normalized
magnetoresistance, $\Delta R/R$,~vs. $B$, is shown at the
different $T$ cross-sections indicated in the top panel. (c) This
panel compares the $T$-dependence of the normalized resistance at
the magnetic field cross-sections at $B=0$ mT, $B=60$ mT, and
$B=~120$ mT. }}
\end{figure}

A color plot of normalized magnetoresistance for a third $Bi_2Te_3$
specimen, sample $3$, is shown in Fig. 3(a). The data corresponding
to the horizontal-cross sections are shown in Fig. 3(b). Here,
$\Delta R/R$  shows predominantly negative magnetoresistance at the
highest temperatures. Fig. 3 (b) also shows a small sharp dip in
$\Delta R/R$ in the $T=3.3$ K data trace in the vicinity of $B=0$.
The dip quickly reaches its maximum depth near $T = 2.9$ K. With a
further reduction in $T$, the resistance correction rapidly
increases in width. The constant magnetic field cross sections indicated
in Fig. 3(a) are shown in Fig. 3(c). As for specimens 1 and 2, the
$\Delta R/R$ drops within a small $\Delta T$ interval and then
plateaus.

\begin{figure}[t]
\centering
\includegraphics[width=6.5cm]{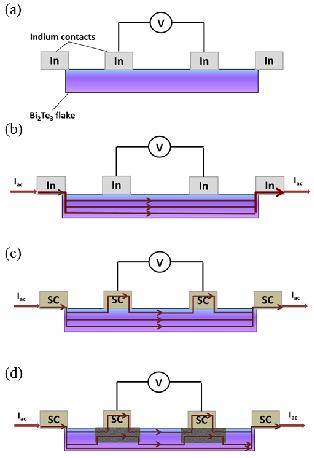}
\caption{\label{fig:Fig4} {\textbf{Possible scenario for observed effects.} A traditional proximity effect based explanation for the observed effects: (a) A sketch of a
Bi$_2$Te$_3$ flake with superconducting indium contacts. (b) A Bi$_2$Te$_3$ flake with a bulk current in the normal
state above the contact  superconducting transition critical
temperature, $T_{c1}$. (c) Below the $T_{c1}$, some additional fraction of the
current $I$ is shunted through the contacts when the contacts are
in the superconducting state. (d) Below the critical temperature,
$T_{c2}$, associated with the proximity effect, Cooper pairs on the TI side of the interface provide an additional shunt for the current.
}}
\end{figure}

This magnetotransport study of $Bi_2Te_3$ with indium contacts shows
the following features: (a) Above the superconducting transition
temperature of indium, the specimens show either a weak positive
(Fig. 1 (b)), mixed type  (Fig. 2(b)), or weak negative (Fig. 3(b))
magnetoresistance. This aspect will be examined in greater detail
elsewhere. At the moment, we attribute the observed differences in
the normal state magnetoresistance characteristics to variations in
the concentration of tellurium vacancies in the bulk crystals
utilized to realize these specimens. (b) Below the transition
temperature of indium, $T_{c} = 3.41$ K,\cite{SuperconductivityTc},
there is a sharp drop in the resistance in the vicinity of $B=0$,
which grows rapidly with decreasing temperatures down to $T \approx
2.9$ K. The critical magnetic field associated with this resistance
correction is $B_{c1} = \Delta B_1/2 < B_{c}^{In}(0~$K$)$, consistent
with the known critical field for indium, $B_{c}^{In}(0~$K$) =
0.0281$ T.\cite{Tinkham} (c). Below $T \approx 3.1$ K, there appears
an additional correction in $\Delta R/R$ which mainly affects the
width of the resistance anomaly with the decreasing temperatures. The
critical magnetic field associated with this term $B_{c2} = \Delta
B_2/2 \le 0.2$ T easily exceeds the critical field of indium. (d)
Constant magnetic field cross sections show that the drop in the
resistance occurs over a narrow temperature range (see Fig. 1(c),
2(c), and 3(c)), and this is followed by an apparent plateau in the
temperature dependence of $\Delta R/R$. Due to the correlation of
the observed effects with superconductivity in the contacts, one
may, at the outset, rule out weak localization and
electron-interactions as the origin of the observed
magnetoresistance anomalies.\cite{Bergmann, Liu2012, Mani2013,
Mani1991}

When a normal metal (N) is brought into good contact with a
superconductor (S), Cooper pairs from the superconductor can leak
into the normal metal and help to manifest signs of
superconductivity in N, while weakening the superconductivity in
S.\cite{Blonder1982} The spatial extent of the proximity effect in N
depends also on the presence of impurities and disorder dependent
dephasing mechanisms in N.\cite{Klapwijk2004} Thus, Andreev
reflection theory encompasses both the ideal ballistic N-S
interfaces, as well as  real-world N-S systems in the diffusive
limit (see, for example, Ref.\cite{Gueron1996}).

One possible image for the observed effects is presented in Fig.
4. The voltage and current probes (see Fig. 4(a)) used in these
experiments consist of the s-wave superconductor indium. Fig. 4(b)
illustrates the four terminal transport measurement at $T > T_{c1}$.
In this high temperature condition, the voltage probes behave as
normal metals. Further, the applied current is transported by the
surface states of the TI as well as by the bulk of $Bi_2Te_3$. When
the temperature is reduced to $T < T_{c1}$, the current and voltage
probes become superconducting (see Fig. 4(c)), and some additional
 current next to the surface is shunted through the contacts with the onset of superconductivity in indium, and  the contact shunt current path is favored over the
current path within the TI. This effect explains the resistance drop
associated with $\Delta B_1$ in the $\Delta R/R$ data. A further
decrease in the temperature could result in a  leakage of Cooper
pairs into the topological insulator (see Fig. 4(d)), which can be
characterized by a critical temperature $T_{c2} \le T_{c1}$. If this
proximity superconductivity layer shunts additional current, there
will be another correction in the resistance, and this could provide
for the observed broad resistance correction at the lowest
temperatures in our specimens.

There is, however,  the observable feature  that the critical field
$B_{c2} = \Delta B_2/2$ is as large as 0.2 T at the lowest
temperatures. This $B_{c2}$ exceeds  by nearly a factor-of-ten the
critical field of indium, $B_{c}^{In} (0~$K$)$. From our understanding,
it is not possible to have a proximity effect in the neighboring
normal metal above the critical field of the superconductor.

Note, however, the reported result that $Bi_2Te_3$ exhibits
superconductivity under the application of pressure in the range of
3--6 GPa.\cite{Hasan2010} Further, Ref.\cite{Zhang2011} indicates that induced
superconductivity in $Bi_2Te_3$ tends to exhibit a $T_c \approx 3$ K
and $B_c \approx 0.2$ T.\cite{Zhang2011} Remarkably, our data
also indicate a $T_{c2} \approx 3.1$ K and $B_{c2} \le 0.2 $ T.
This point suggests that  the second critical transition associated
with $B_{c2}$ could originate from a contact induced surface
modification in the $Bi_2Te_3$ flake, which mimics the effect of
pressure, but only in the immediate area under the contact.

\pagebreak
\pagebreak

\section {acknowledgements}
Z.W., T.Y., and  magnetotransport studies in Georgia State University have
been supported by the U.S. Department of Energy, Office of Basic
Energy Sciences, Material Sciences and Engineering Division under
DE-SC0001762. Additional support has been provided by the ARO under
W911NF-14-2-0076.

\section{Author contributions}
Measurements were done by Z.W. Technical assistance was given by T.Y. Experimental
development and manuscript was written by Z.W. and R.G.M. The authors declare no competing financial interests.

\end{document}